\documentclass[a4paper,11pt]{article}
\usepackage{pos}
\usepackage{orcidlink}

\title{$B\rightarrow K + \mbox{\textbf{\textit{invisible}}}$ in a model with axion-like particles}

\author*[\orcidlink{0000-0002-1361-4736}a]{Xiyuan Gao}

\affiliation[a]{Institute for Theoretical Particle Physics, Karlsruhe Institute of Technology (KIT),\\
  Wolfgang-Gaede-Straße 1, D-76131 Karlsruhe, Germany}

\emailAdd{xiyuan.gao@kit.edu}

\abstract{
The localized excess of $B\rightarrow K + \textit{invisible}$ events reported by Belle-II is commonly interpreted as a signal for $B\to Ka$, where $a$ is an axion-like particle (ALP). In these proceedings, we summarize two theoretical updates regarding the $b\to sa$ decay amplitude within a minimal UV-complete model for invisible ALPs, namely the DFSZ model. (i) Contributions from certain two-loop diagrams can dominate the one-loop ones and are thus relevant for phenomenology. (ii) Although not unique, a rare feature --- which we refer to as apparent non-decoupling --- emerges in the light effective theory. The renormalizable ALP interactions fail to capture this behavior and therefore are incomplete as low-energy effective descriptions. 
}

\FullConference{The European Physical Society Conference on High Energy Physics (EPS-HEP2025)\\
7-11 July 2025\\
Marseille, France\\}

\begin{document}
\maketitle

\section{Introduction}

In late 2023, Belle II reported an excess in $B\to K \nu\overline{\nu}$ over the Standard Model (SM) prediction~\cite{Belle-II:2023esi}. The branching fraction obtained using the more recent Inclusive Tagging Analysis (ITA) method shows a 2.9 standard deviation departure. A combination of both conventional Hadronic Tagging Analysis (HTA) method and ITA method yields $\text{Br}(B^+\to K^+\nu\overline{\nu})=(2.3\pm 0.7)\times 10^{-5}$, showing a 2.7 standard deviation departure. The excess is rather localized in the visible kaon energy. The authors of Ref.~\cite{Altmannshofer:2023hkn} find the two-body decay $B^+\to K^+ a$ gives excellent fit to the data with a significance of $3.6$ standard deviation. Here, $a$ is an invisible particle with $m_a\approx 2$ GeV.

Although the data hints at a new particle, its lightness is puzzling from the theory side. Similar to the SM Higgs boson, the masses of scalar particles are in general fine-tuned and one has to bear implicit naturalness. Therefore, $a$ is commonly interpreted as a pesudo-Goldstone boson of a spontaneously broken global symmetry $G$~\cite{Batell:2009jf, Izaguirre:2016dfi, Aloni:2018vki,Chakraborty:2021wda, Gavela:2019wzg, Bauer:2021mvw, Bisht:2024hbs, Calibbi:2025rpx, MartinCamalich:2025srw, Alda:2025uwo, Abumusabh:2025zsr}, which remains naturally massless when $G$ is exact. If $G$ is a chiral symmetry, for instance, the Peccei-Quinn $U(1)_{\text{PQ}}$~\cite{PQ1, PQ2},  $a$ is referred to as an axion-like particle (ALP). Unlike a singlet scalar particle which could directly mix with the physical SM Higgs boson~\cite{Kachanovich:2020yhi}, one can not add $a$ to the SM via renormalizable interactions. Model building is needed and one minimal benchmark is the so-called DFSZ model~\cite{DFSZ1, DFSZ2}. It extends the SM by one additional Higgs doublet and one complex scalar. The extended scalar and Yukawa sectors read:
\begin{equation}
\begin{aligned}
\label{DFSZpotential}
    V_{\Phi}~=&~ \Tilde{V}_{\text{moduli}}(|\Phi_u|,|\Phi_d|,|\Phi_u \Phi_d|,|\Phi_s|)+\lambda \Phi_s^{2} \Phi_u^{~} \Phi_d^{\dagger} +\text{H.c.},\\
    \mathcal{L}_{Y}~=&~ Y_u \overline{Q_L}u_R \Phi_u +Y_d \overline{Q_L}d_R \widetilde{\Phi}_d+Y_e \overline{L_L}e_R \widetilde{\Phi}_u+\text{H.c.},
\end{aligned}
\end{equation}
where $\Phi_{u,d}$ are $SU(2)$ doublets and $\Phi_s$ is singlet and $\widetilde{\Phi}_d=i\sigma_2 \Phi_d^{\dagger}$. $\Tilde{V}_{\text{moduli}}$ contains all gauge invariant interactions without phase dependence.


In these proceedings, we summarize our recent work revisiting the DFSZ model~\cite{Gao:2025ohi}.
In Section~\ref{twoloop}, we present a newly identified two-loop enhancement to the $b\to s a$ decay amplitude. Section~\ref{nondecouple} then discusses an apparent non-decoupling behavior and clarifies its implications on different effective theory basis. The predictions and conclusions are provided in Section~\ref{disc}.

\section{Two-loop enhancement}
\label{twoloop}

The physical spectrum of the DFSZ model in Eq.~(\ref{DFSZpotential}) contains two pesudo-scalars --- $a_0$ as the axial part of $\Phi_s$ and $A_0$ which is a combination of $\Phi_u$ and $\Phi_d$. Ref.~\cite{Freytsis:2009ct} pointed out that the $b\to s a$ amplitude can be calculated by introducing an $a_0-A_0$ mixing angle $\theta$:
\begin{equation}
    \label{LigetiResult}
    \mathcal{A}(b\rightarrow s a)_{\text{DFSZ}}~=~-\sin{\theta}\times \mathcal{A}(b\rightarrow s A_0)_{\text{PQWW}},
\end{equation}
Here, $\mathcal{A}(b\rightarrow s A_0)_{\text{PQWW}}$ is calculated in the PQWW model~\cite{PQ1, PQ2, WW1, WW2} where $A_0$ is also massless~\cite{Wise:1980ux, Hall:1981bc, Frere:1981cc}. However, $A_0$ becomes massive in the DFSZ model so that the off-shell $b\rightarrow s A_0$ amplitude is unphysical and cannot be uniquely defined. Besides, $a$ can interact with the Higgs doublets with the coupling strength proportional to $\lambda$, which vanishes in the PQWW model but not in DFSZ. In other words:
\begin{equation}
\begin{aligned}
    \mathcal{A}(b\rightarrow s a)_{\text{DFSZ}}\neq-\sin{\theta}\times\mathcal{A}(b\rightarrow s A_0)_{\text{DFSZ}},~~~
    \mathcal{A}(b\rightarrow s A_0)_{\text{DFSZ}}\neq\mathcal{A}(b\rightarrow s A_0)_{\text{PQWW}}.
\end{aligned}
\end{equation}
These two new effects are both proportional to $m_{A_0}^2$ and at one loop order, we find they cancel each other even if $A_0$ is heavy. Eq.~(\ref{LigetiResult})~claimed in Ref.~\cite{Freytsis:2009ct}~is therefore correct.

Eq.~(\ref{LigetiResult}) does not necessarily hold at a higher loop order. The DFSZ model shown in Eq.~(\ref{DFSZpotential}) also admits an $Z_2$ symmetry:
\begin{equation}
    \label{Z2sym}
    d_R\rightarrow-d_R, \quad \Phi_d\rightarrow-\Phi_d, \quad  \text{others unchanged,}
\end{equation}
which can be broken by a non-zero $\lambda$ and/or the vacuum expectation value (VEV) of $\Phi_d$. This fixes the effective Hamiltonian for $b\to s a$ decay up to loop factors: 
\begin{equation}
\label{Hamilton}
\begin{aligned}
    \mathcal{H}_{\text{eff}}~=&~ \theta g^3 \frac{ V_{ts}^*V_{tb}}{128\pi^2} \frac{m_t^2}{m_W^3}
    \times
    \left(
    X_1\frac{1}{\tan\beta}+X_2\frac{1}{\tan^3\beta}+X_3\frac{\lambda}{16\pi^2}
    \right) \overline{s}\gamma^{\mu}P_Lb~\partial_{\mu} a.
\end{aligned}
\end{equation}
Here, $\theta\equiv \frac{2 \langle\Phi_d\rangle}{\langle \Phi_s \rangle},~\frac{1}{\tan\beta}\equiv \frac{\langle \Phi_d\rangle}{\langle \Phi_u \rangle},$ and $\lambda$ are $Z_2$ spurions. $\lambda$ enters the expression together with the loop factor $1/16\pi^2$ because it carries Planck Units $\hbar$ and has to arise at the next order of the perturbative expansion. $X_1, X_2$ are provided in Ref.~\cite{Freytsis:2009ct} and \cite{Hall:1981bc}, while $X_3$ is what we recently found:
\begin{equation}
\label{X1X2X3}
    X_3~=~\log{\frac{m_{H}^2}{m_t^2}}+\frac{6m_W^2}{m_t^2-m_W^2}\log{\frac{m_t^2}{m_W^2}}+\frac{1}{2}.
\end{equation}
See Ref.~\cite{Gao:2025ohi} for more technical details. We only give the two following comments here:
\begin{enumerate}
    \item Gauge independence: We have worked with the $R_{\xi}$ gauge. By showing that $X_3$ does not depend on $\xi$, we have cross-checked our result.
    \item $\tan\beta$ enhancement: The $X_3$ contribution is not always sub-leading because the $X_1$ and $X_2$ terms are suppressed when $\tan\beta$ is sizable. 
\end{enumerate}
We illustrated how such two-loop enhancement changes phenomenology in Fig. 3. of Ref.~\cite{Gao:2025ohi}.

\section{Apparent Non-decoupling}
\label{nondecouple}

In Eq.~(\ref{X1X2X3}), $m_H\equiv m_{H^+}\simeq m_{A_0},$ is defined as a heavy mass scale. The loop contribution $X_3$ does not vanish in the limit $m_H\to \infty$ but diverges logarithmically. Similar behavior is also reported for the $X_1$ term~\cite{Freytsis:2009ct}, and only $X_2$ vanishes in the large $m_H$ limit. Decoupling is hidden in the mixing angle $\theta\sim m_H^{-1}$. Given finite $\theta$, $m_H$ cannot assume arbitrarily large values. This is uncommon but not unique. For instance, the misalignment parameter $c_{\alpha\beta}$ in the 2HDM is similarly proportional to $m_H^{-2}$~\cite{Gunion:2002zf}. Part of the induced $\mu\to e\gamma$ amplitude is not directly suppressed by $m_H^{-2}$ but indirectly by $c_{\alpha\beta}$. Apparently, such behaviors challenge the Wilsonian decoupling picture~\cite{Wilson:1974mb}, because the dimensionless couplings inherited with $\theta$ or $c_{\alpha\beta}$ can not take the `naively expected' $\mathcal{O}(1)$ values. We understand as the reason for this that without the heavy degrees of freedom, the light theory is not invariant under $SU(2)_L\times U(1)_Y$. Gauge invariance, a necessary condition for decoupling~\cite{Senjanovic:1979yq}, is not satisfied without further UV completion.

This apparent non-decoupling feature tells that the $H^+$ and $W^+$ contributions to $b\to s a$ are equally important in the DFSZ model. We find the enhancement can be explicitly identified with the $G^-H^+a$ vertex, whose dimensionful coupling strength is $g\theta \frac{m_H^2}{m_W}$. It cancels the $m_H^2$ term contained in the denominator of the Feynman integral so the amplitude is suppressed by $\theta$ only. Consequently, if isolating the heavy $H^+$ interactions in the DFSZ model, the light theory, although seemingly renormalizable, cannot reproduce the full $b\to sa$ amplitude. $H^+$ must be integrated out and the complete low-energy theory should contain an effective vertex involving both $a$ and $G^-$. Such arguments motivate us to check the low-energy basis for ALP interactions.

The low-energy theory respects the QED$\times$QCD symmetry only and the quarks and leptons are vector-like. The ALP $a$ can couple to quarks with renormalizable interactions, for instance: 
\begin{equation}
    \label{lightsubtheory}
    \mathcal{L}~=~ \mathcal{L}_{\text{SM}}~+~ i a \sum_{q=t,b} c_{q}~\overline{q} \gamma_5 q. 
\end{equation}
If $c_q$ is matched from DFSZ model, Eq.~(\ref{lightsubtheory}) with the renormalizable $R_{\xi}$ gauge can not reproduce the logarithmic term in $X_1$. The correct theory must contain the non-decoupling effect of $H^+$, by either adding more renormalizable interactions or effective operators containing $a$ and $G^{\pm}$. The second choice scarifies renormalizability but restores the $SU(2)_L\times U(1)_Y$ gauge invariance. The correct low energy interacting basis read: 
\begin{equation}
\begin{aligned}
    \label{EFTS2}
    \mathcal{L}
     ~=&~\mathcal{L}_{\text{SM}}+i \frac{a}{v}\left( c_b \overline{Q}_L  b_R \widetilde{H}_u+c_t ~\overline{Q}_Lt_RH_u+\text{H.c.}  \right) \\
    =&~\mathcal{L}_{\text{SM}}+i a \sum_{q=t,b} c_{q}\overline{q} \gamma_5 q
    + i \frac{a}{v}\left[ c_b V_{tb}^{}  \overline{t}_L  b_R G^++c_t \left(V_{tb}^{*}\overline{b}_L + V_{ts}^{*}\overline{s}_L \right)t_R G^-+\text{H.c.}  \right]+...\\
    =&~\mathcal{L}_{\text{SM}}+\sum_{\psi_L=Q_L,t_R^c, b_R^c}\frac{c_{\psi}}{f}~\overline{\psi}_L\gamma^{\mu} \psi_L ~\partial_{\mu}a+...
\end{aligned}
\end{equation}
Integration by part and classical equation of motion is applied in the last step and we drop the irrelevant higher order terms. We checked that the dimension-5 operators containing $G^{\pm}$ reproduce the missing piece for the logarithmic term. Ref.~\cite{Dolan:2014ska}, the authors reported a discrepancy by a factor of 4 between the pesudo-scalar basis and the derivative basis, and connected it to the Higgs boson involved in the dimension-5 operators. Here, we emphasize that Eq.~(\ref{EFTS2}) is the consistent low energy basis for ALP interaction while Eq.~(\ref{lightsubtheory}) with $R_{\xi}$ gauge is not.

Eq.~(\ref{EFTS2}) reduces to Eq.~(\ref{lightsubtheory}) with the gauge fixing condition $G^{\pm}(x^{\mu})=0$. As a consequence, Eq.~(\ref{lightsubtheory}) works only with the unitary (non-normalizable) gauge, where the $SU(2)_L$ gauge symmetry is hidden but still preserved. We have explicitly checked the that the missing piece for the logarithmic term can also arise from the longitudinal part of the $W$ propagator $k_{\mu}k_{\nu}/m_W^2$.

Although this point is often implicitly assumed, we consider it also important to emphasize here that the effective theory can only determine the logarithmic part of $X_1$ through predicting the anomalous dimensions. Beyond the terms that diverge as $m_H\to \infty$, the finite pieces are not physical predictions; rather, they are definitions for the renormalization scheme, specifying how counter terms cancel the divergences.

\section{Discussions}
\label{disc}

In these proceedings, we revisit the DFSZ model prediction for the $b\to sa$ decay amplitude. We find new two-loop contribution which can be enhanced by sizable $\tan\beta$ and identify a low-energy basis inconsistent with the UV theory. 
The model is motivated by a localized Belle II excess in $B\to K\nu\overline{\nu}$ over the SM prediction, which can be interpreted as a signal of $B\to K a$. Since this decay amplitude is suppressed by the CKM factor $V_{ts}$ and the loop factor $1/(16\pi^2)$, the underlying particles of the UV completion are not necessarily extremely heavy and may induce a sizable $\Upsilon\to \gamma+a$ decay rate. To ensure that $a$ escapes detection, the DSFZ model must be extended with an invisible sector, which suppresses the visible $a$ decay branching ratios. Rare visible processes such as $B\to K a\to K\mu^+\mu^-$ can also emerge as potential signatures of the model.

\begin{acknowledgments}

The author thanks Ulrich Nierste and Robert Ziegler for useful discussions and comments on the manuscript. This research was supported by the Deutsche Forschungsgemeinschaft (DFG, German Research Foundation) under grant 396021762 - TRR 257 and by the BMBF Grant 05H21VKKBA, \textit{Theoretische Studien für Belle II und LHCb.} X.G. also acknowledges the support by the Doctoral School ``Karlsruhe School of Elementary and Astroparticle Physics: Science and Technology.''

\end{acknowledgments}

\bibliographystyle{JHEP}
\bibliography{2HDMa.bib}

\providecommand{\noopsort}[1]{}\providecommand{\singleletter}[1]{#1}%

\providecommand{\href}[2]{#2}\begingroup\raggedright\begin{thebibliography}{10}

\bibitem{Belle-II:2023esi}
{\scshape Belle-II} collaboration, I.~Adachi et~al., \emph{{Evidence for $B^+\to K^+\nu \bar \nu$ decays}}, \href{https://doi.org/10.1103/PhysRevD.109.112006}{\emph{Phys. Rev. D} {\bfseries 109} (2024) 112006}, [\href{https://arxiv.org/abs/2311.14647}{{\ttfamily 2311.14647}}].

\bibitem{Altmannshofer:2023hkn}
W.~Altmannshofer, A.~Crivellin, H.~Haigh, G.~Inguglia and J.~Martin~Camalich, \emph{{Light new physics in $B\to K^{(*)}\nu \bar \nu$?}}, \href{https://doi.org/10.1103/PhysRevD.109.075008}{\emph{Phys. Rev. D} {\bfseries 109} (2024) 075008}, [\href{https://arxiv.org/abs/2311.14629}{{\ttfamily 2311.14629}}].

\bibitem{Batell:2009jf}
B.~Batell, M.~Pospelov and A.~Ritz, \emph{{Multi-lepton Signatures of a Hidden Sector in Rare B Decays}}, \href{https://doi.org/10.1103/PhysRevD.83.054005}{\emph{Phys. Rev. D} {\bfseries 83} (2011) 054005}, [\href{https://arxiv.org/abs/0911.4938}{{\ttfamily 0911.4938}}].

\bibitem{Izaguirre:2016dfi}
E.~Izaguirre, T.~Lin and B.~Shuve, \emph{{Searching for Axionlike Particles in Flavor-Changing Neutral Current Processes}}, \href{https://doi.org/10.1103/PhysRevLett.118.111802}{\emph{Phys. Rev. Lett.} {\bfseries 118} (2017) 111802}, [\href{https://arxiv.org/abs/1611.09355}{{\ttfamily 1611.09355}}].

\bibitem{Aloni:2018vki}
D.~Aloni, Y.~Soreq and M.~Williams, \emph{{Coupling QCD-Scale Axionlike Particles to Gluons}}, \href{https://doi.org/10.1103/PhysRevLett.123.031803}{\emph{Phys. Rev. Lett.} {\bfseries 123} (2019) 031803}, [\href{https://arxiv.org/abs/1811.03474}{{\ttfamily 1811.03474}}].

\bibitem{Chakraborty:2021wda}
S.~Chakraborty, M.~Kraus, V.~Loladze, T.~Okui and K.~Tobioka, \emph{{Heavy QCD axion in b\textrightarrow{}s transition: Enhanced limits and projections}}, \href{https://doi.org/10.1103/PhysRevD.104.055036}{\emph{Phys. Rev. D} {\bfseries 104} (2021) 055036}, [\href{https://arxiv.org/abs/2102.04474}{{\ttfamily 2102.04474}}].

\bibitem{Gavela:2019wzg}
M.~B. Gavela, R.~Houtz, P.~Quilez, R.~Del~Rey and O.~Sumensari, \emph{{Flavor constraints on electroweak ALP couplings}}, \href{https://doi.org/10.1140/epjc/s10052-019-6889-y}{\emph{Eur. Phys. J. C} {\bfseries 79} (2019) 369}, [\href{https://arxiv.org/abs/1901.02031}{{\ttfamily 1901.02031}}].

\bibitem{Bauer:2021mvw}
M.~Bauer, M.~Neubert, S.~Renner, M.~Schnubel and A.~Thamm, \emph{{Flavor probes of axion-like particles}}, \href{https://doi.org/10.1007/JHEP09(2022)056}{\emph{JHEP} {\bfseries 09} (2022) 056}, [\href{https://arxiv.org/abs/2110.10698}{{\ttfamily 2110.10698}}].

\bibitem{Bisht:2024hbs}
D.~Bisht, S.~Chakraborty and A.~Samanta, \emph{{A comprehensive study of ALPs from B-decays}}, \href{https://doi.org/10.1007/JHEP07(2025)092}{\emph{JHEP} {\bfseries 07} (2025) 092}, [\href{https://arxiv.org/abs/2412.09678}{{\ttfamily 2412.09678}}].

\bibitem{Calibbi:2025rpx}
L.~Calibbi, T.~Li, L.~Mukherjee and M.~A. Schmidt, \emph{{Is dark matter the origin of the $B\to K \nu\bar\nu$ excess at Belle II?}}, \href{https://doi.org/10.1103/r2gw-rwzw}{\emph{Phys. Rev. D} {\bfseries 112} (2025) 075020}, [\href{https://arxiv.org/abs/2502.04900}{{\ttfamily 2502.04900}}].

\bibitem{MartinCamalich:2025srw}
J.~Martin~Camalich and R.~Ziegler, \emph{{Flavor Phenomenology of Light Dark Sectors}}, \href{https://doi.org/10.1146/annurev-nucl-121423-100931}{\emph{Ann. Rev. Nucl. Part. Sci.} {\bfseries 75} (2025) 223--246}, [\href{https://arxiv.org/abs/2503.17323}{{\ttfamily 2503.17323}}].

\bibitem{Alda:2025uwo}
J.~Alda, M.~Fuentes~Zamoro, L.~Merlo, X.~Ponce~D{\'\i}az and S.~Rigolin, \emph{{Comprehensive ALP Searches in Meson Decays}},  \href{https://arxiv.org/abs/2507.19578}{{\ttfamily 2507.19578}}.

\bibitem{Abumusabh:2025zsr}
M.~Abumusabh, G.~Dujany, D.~Guadagnoli, A.~Iohner and C.~Toni, \emph{{The $B^+ \to K^+ \nu\bar \nu$ decay as a search for the QCD axion}},  \href{https://arxiv.org/abs/2510.18953}{{\ttfamily 2510.18953}}.

\bibitem{PQ1}
R.~D. Peccei and H.~R. Quinn, \emph{{CP Conservation in the Presence of Instantons}}, \href{https://doi.org/10.1103/PhysRevLett.38.1440}{\emph{Phys. Rev. Lett.} {\bfseries 38} (1977) 1440--1443}.

\bibitem{PQ2}
R.~D. Peccei and H.~R. Quinn, \emph{{Constraints Imposed by CP Conservation in the Presence of Instantons}}, \href{https://doi.org/10.1103/PhysRevD.16.1791}{\emph{Phys. Rev.} {\bfseries D16} (1977) 1791--1797}.

\bibitem{Kachanovich:2020yhi}
A.~Kachanovich, U.~Nierste and I.~Ni\v{s}and\v{z}i\'c, \emph{{Higgs portal to dark matter and $B\to K^{(*)}$ decays}}, \href{https://doi.org/10.1140/epjc/s10052-020-8240-z}{\emph{Eur. Phys. J. C} {\bfseries 80} (2020) 669}, [\href{https://arxiv.org/abs/2003.01788}{{\ttfamily 2003.01788}}].

\bibitem{DFSZ1}
M.~Dine, W.~Fischler and M.~Srednicki, \emph{{A Simple Solution to the Strong CP Problem with a Harmless Axion}}, \href{https://doi.org/10.1016/0370-2693(81)90590-6}{\emph{Phys. Lett.} {\bfseries B104} (1981) 199--202}.

\bibitem{DFSZ2}
A.~R. Zhitnitsky, \emph{{On Possible Suppression of the Axion Hadron Interactions. (In Russian)}}, {\emph{Sov. J. Nucl. Phys.} {\bfseries 31} (1980) 260}.

\bibitem{Gao:2025ohi}
X.~Gao and U.~Nierste, \emph{{$B\to K+ \text{axionlike particles}$: Effective versus UV-complete models and enhanced two-loop contributions}}, \href{https://doi.org/10.1103/5j2t-2kdf}{\emph{Phys. Rev. D} {\bfseries 112} (2025) 055008}, [\href{https://arxiv.org/abs/2506.14876}{{\ttfamily 2506.14876}}].

\bibitem{Freytsis:2009ct}
M.~Freytsis, Z.~Ligeti and J.~Thaler, \emph{{Constraining the Axion Portal with $B \to K l^+ l^-$}}, \href{https://doi.org/10.1103/PhysRevD.81.034001}{\emph{Phys. Rev. D} {\bfseries 81} (2010) 034001}, [\href{https://arxiv.org/abs/0911.5355}{{\ttfamily 0911.5355}}].

\bibitem{WW1}
S.~Weinberg, \emph{{A New Light Boson?}}, \href{https://doi.org/10.1103/PhysRevLett.40.223}{\emph{Phys. Rev. Lett.} {\bfseries 40} (1978) 223--226}.

\bibitem{WW2}
F.~Wilczek, \emph{{Problem of Strong p and t Invariance in the Presence of Instantons}}, \href{https://doi.org/10.1103/PhysRevLett.40.279}{\emph{Phys. Rev. Lett.} {\bfseries 40} (1978) 279--282}.

\bibitem{Wise:1980ux}
M.~B. Wise, \emph{{Radiatively induced flavor changing neutral Higgs boson couplings}}, \href{https://doi.org/10.1016/0370-2693(81)90684-5}{\emph{Phys. Lett. B} {\bfseries 103} (1981) 121--123}.

\bibitem{Hall:1981bc}
L.~J. Hall and M.~B. Wise, \emph{{Flavor-changing Higgs boson couplings}}, \href{https://doi.org/10.1016/0550-3213(81)90469-7}{\emph{Nucl. Phys. B} {\bfseries 187} (1981) 397--408}.

\bibitem{Frere:1981cc}
J.~M. Frere, J.~A.~M. Vermaseren and M.~B. Gavela, \emph{{The Elusive Axion}}, \href{https://doi.org/10.1016/0370-2693(81)90686-9}{\emph{Phys. Lett. B} {\bfseries 103} (1981) 129--133}.

\bibitem{Gunion:2002zf}
J.~F. Gunion and H.~E. Haber, \emph{{The CP conserving two Higgs doublet model: The Approach to the decoupling limit}}, \href{https://doi.org/10.1103/PhysRevD.67.075019}{\emph{Phys. Rev. D} {\bfseries 67} (2003) 075019}, [\href{https://arxiv.org/abs/hep-ph/0207010}{{\ttfamily hep-ph/0207010}}].

\bibitem{Wilson:1974mb}
K.~G. Wilson, \emph{{The Renormalization Group: Critical Phenomena and the Kondo Problem}}, \href{https://doi.org/10.1103/RevModPhys.47.773}{\emph{Rev. Mod. Phys.} {\bfseries 47} (1975) 773}.

\bibitem{Senjanovic:1979yq}
G.~Senjanovic and A.~Sokorac, \emph{{On the Decoupling of Superheavy Particles at Low-energies}}, \href{https://doi.org/10.1016/0550-3213(80)90513-1}{\emph{Nucl. Phys. B} {\bfseries 164} (1980) 305--332}.

\bibitem{Dolan:2014ska}
M.~J. Dolan, F.~Kahlhoefer, C.~McCabe and K.~Schmidt-Hoberg, \emph{{A taste of dark matter: Flavour constraints on pseudoscalar mediators}}, \href{https://doi.org/10.1007/JHEP03(2015)171}{\emph{JHEP} {\bfseries 03} (2015) 171}, [\href{https://arxiv.org/abs/1412.5174}{{\ttfamily 1412.5174}}].

\end{thebibliography}\endgroup

\end{document}